\documentclass[journal=jctcce,manuscript=article,layout=twocolumn]{achemso}

\usepackage[version=3]{mhchem} % Formula subscripts using \ce{}

\makeatletter
\let\l@addto@macro\relax
\makeatother
\usepackage[fontsize=10pt]{scrextend}

\usepackage{graphicx}
\usepackage{dcolumn}
\usepackage{bm}
\usepackage{amssymb}
\usepackage{mdframed}
\usepackage{newfloat}
\DeclareFloatingEnvironment[
    fileext=los,
    listname={Code Listings},
    name=Listing,
    placement=tbhp,
    within=none,
]{listing}

\usepackage[frozencache,cachedir=.]{minted2}
\usepackage{pifont}
\usepackage{wasysym}
\usepackage{newunicodechar}

\newunicodechar{●}{$\CIRCLE$}
\newunicodechar{○}{$\Circle$}
\newunicodechar{➊}{\ding{202}}
\newunicodechar{➋}{\ding{203}}
\newunicodechar{²}{$^2$}
\newunicodechar{⁻}{$^-$}
\newunicodechar{⨉}{$\times$}
\newunicodechar{₅}{$_\texttt{5}$}

\usepackage{bookmark}
\usepackage{setspace}

\usepackage{xcolor}
\definecolor{codeblue}{rgb}{0,0.3,0.6}

\NewDocumentCommand{\codeword}{v}{%
\texttt{#1}%
}

\newcommand{\codewordcaption}[1]{\texttt{#1}}
\newcommand{\textblue}[1]{{#1}}

\usepackage{xcolor}

\usepackage{enumitem}
\setlist[itemize]{noitemsep, topsep=0pt}
\setlist[enumerate]{noitemsep, topsep=0pt}

\usepackage{xspace}
\newcommand{\freebird}{\texttt{FreeBird.jl}\xspace}

\newcommand{\CV}{$C_V$\xspace}

\author{Ray Yang}
\affiliation{
Department of Chemistry and Institute of Materials Science and Engineering, Washington University in St.~Louis, St.~Louis, MO 63130, USA
}
\alsoaffiliation{Department of Computer Science and Engineering, McKelvey School of Engineering, Washington University in St.~Louis, St.~Louis, MO 63130, USA}
\author{Junchi Chen}
\affiliation{
Department of Chemistry and Institute of Materials Science and Engineering, Washington University in St.~Louis, St.~Louis, MO 63130, USA
}
\author{Douglas Thibodeaux} 
\affiliation{
Department of Chemistry and Institute of Materials Science and Engineering, Washington University in St.~Louis, St.~Louis, MO 63130, USA
}
\author{Robert B. Wexler}
\email{wexler@wustl.edu}
\affiliation{ 
Department of Chemistry and Institute of Materials Science and Engineering, Washington University in St.~Louis, St.~Louis, MO 63130, USA
}
\title{FreeBird.jl: An Extensible Toolbox for Simulating Interfacial Phase Equilibria}

\begin{document}

% \begin{tocentry}

% Some journals require a graphical entry for the Table of Contents.
% This should be laid out ``print ready'' so that the sizing of the

% Inside the \texttt{tocentry} environment, the font used is Helvetica
% 8\,pt, as required by \emph{Journal of the American Chemical

% The surrounding frame is 9\,cm by 3.5\,cm, which is the maximum
% permitted for  \emph{Journal of the American Chemical Society}
% graphical table of content entries. The box will not resize if the
% content is too big: instead it will overflow the edge of the box.

% This box and the associated title will always be printed on a
% separate page at the end of the document.

% \includegraphics[width=1\textwidth]{FreeBird.pdf}

% \end{tocentry}

\begin{abstract}
We present \freebird, an extensible Julia-based platform for computational studies of phase equilibria at generic interfaces.
The package supports a range of system configurations, from atomistic solid surfaces to coarse-grained lattice--gas models, with energies evaluated using classical interatomic potentials or lattice Hamiltonians.
Both atomistic and lattice systems accommodate single- or multi-component mixtures with flexibly definable surface and lattice geometries.
Implemented sampling algorithms include nested sampling, Wang--Landau sampling, Metropolis Monte Carlo, and, for tractable lattice systems, exact enumeration.
Leveraging Julia's type hierarchies and multiple dispatch, \freebird provides a modular interface that allows seamless integration of system definitions, energy evaluators, and sampling schemes.
Designed for flexibility, extensibility, and performance, \freebird offers a versatile framework for exploring the thermodynamics of interfacial phenomena.
\end{abstract}


\section{\label{sec:introduction}Introduction}

Interfaces are central to technologies addressing today's most pressing challenges, from sustainable heterogeneous catalysis \cite{soderstedt_oxidized_2025} to high-performance semiconductor devices \cite{liu_probing_2025} and biomedical innovation. \cite{souza_exploring_2025}
Meeting these challenges requires a multiscale understanding of interfacial processes that can bridge fundamental insights to practical design rules and, ultimately, to commercial technologies. \cite{somorjai_low_1973, somjit_atomic_2022}
At atomic length scales, methods such as low-energy electron diffraction \cite{may_discovery_1970} and advanced electron microscopy \cite{hutner_stoichiometric_2024} can resolve solid--vacuum interfaces with sub-\AA{} precision.
However, these highly sensitive ex situ probes remain difficult to apply under realistic conditions in which the interface is buried beneath a dense phase or driven far from equilibrium, leaving critical gaps in spatial resolution and experimental accessibility. \cite{lander_low-energy_1965, rupprechter_spectroscopic_2008}

Because high-resolution in situ probes remain scarce, \cite{akbashev_probing_2023, schroder_tracking_2023} theoretical and computational approaches have become indispensable for mapping the configurational landscape of interfaces and guiding materials design. \cite{zhang_ensembles_2020, du_machine-learning-accelerated_2023, hormann2025machine, chaudhuri2025challenges}
Modeling an interface under working conditions is essentially a non-equilibrium surface problem whose dynamics depend on the poorly constrained, synthesis-specific starting structure. \cite{phan_emergence_2021, zhou_dynamical_2022}
To address this, researchers typically begin by establishing an equilibrium thermodynamic baseline, \cite{reuter_composition_2001, wexler_automatic_2019} which entails a series of decisions: the level of energy evaluation (first-principles versus classical force fields) \cite{rosenbrock_machine-learned_2021, vandermause_active_2022} and the depth of configurational sampling required to obtain entropies and free energies (harmonic, quasi-harmonic, or fully anharmonic treatments; atomistic versus lattice representations). \cite{campbell_equilibrium_2016, partay_nested_2021, ashton_nested_2022, saunders2025comprehensive}

Most realistic surface problems are too complex for exhaustive first-principles treatment, forcing researchers to navigate an accuracy--sampling trade-off. \cite{reuter_composition_2001, reuter_ab_2005}
Ab initio thermodynamics, the workhorse for predicting surface phase diagrams, typically evaluates only tens to thousands of candidate structures within a practical computing budget, with selections often guided by chemical intuition. \cite{reuter_composition_2001, wexler_automatic_2019}
Structure-search algorithms such as particle-swarm, \cite{lu_self-assembled_2014} Bayesian, \cite{ulissi_automated_2016, carr_basc_2016} and genetic \cite{bauer_systematic_2022, han_prediction_2023} optimizers expand the candidate pool but still assume a sharply truncated configuration space in which atoms vibrate harmonically about fixed lattice sites. \cite{campbell_equilibrium_2016, ashton_nested_2022}
This harmonic perspective understates the structural and compositional agility of catalytic surfaces; in many cases, comprehensive sampling with well-parameterized classical potentials can reveal low-energy motifs that sparse, high-accuracy DFT surveys may miss. \cite{rosenbrock_machine-learned_2021, vandermause_active_2022, yang_surface_2024, chatbipho_adsorbate_2025}

Although progress has been made toward systematically improvable thermodynamic sampling of surface phases, \cite{zhou_determining_2019, mambretti_how_2024, yang_surface_2024} the field still lacks a fast, reproducible benchmarking platform dedicated to interfacial structure prediction, applicable to solid--vacuum/gas systems as well as defects at solid--solid and solid--liquid boundaries.
We therefore introduce \freebird, an open-source Julia package featuring a highly parallel, modular architecture that allows users to interchange system resolutions, energy models, and Monte Carlo--style algorithms without rewriting code.
Distinct from existing packages, \cite{van_de_walle_alloy_2002, martiniani_nested_sampling_2013, martiniani_sens_2013, martiniani_mcpele_2014, partay_pymatnest_2015, shah_cassandra_2017, angqvist_icet_2019, barroso-luque_smol_2022, bernstein_pymatnext_2023} it integrates nested sampling, Wang--Landau, Metropolis Monte Carlo, and (for suitable lattice systems) exact enumeration into a single extensible framework.
This framework spans both atomistic and lattice representations with interchangeable energy calculators and move sets, enabling direct, reproducible benchmarking and comparison of interfacial phase equilibria across methods.
\freebird is an active project; planned updates include superposition-enhanced nested sampling, \cite{martiniani_superposition_2014} parallel tempering, \cite{swendsen_replica_1986} transition-matrix Monte Carlo, \cite{wang_transition_1999} geometry-aware coarse-graining (e.g., Gay--Berne \cite{berne_gaussian_1972, gay_modification_1981} and patchy-particle \cite{zhang_self-assembly_2004} models), and full GPU acceleration.
Our long-term vision is to position \freebird at the core of autonomous, multi-fidelity workflows (integrating Bayesian experimental design, interface-aware machine-learning potentials, and a surface-structure database) to accelerate the discovery and design of functional interfaces.

This paper is organized as follows: Section~\ref{sec:imp} outlines the overall structure of the \freebird package, describing its capabilities for handling both continuous atomistic and discretized lattice systems, and briefly reviewing the sampling algorithms it implements.
We then present demonstrations and comparisons using various sampling schemes in \freebird to identify phase transitions in several atomistic Lennard-Jones systems (a pure cluster, a binary cluster, and a surface--adsorbate system) as well as in discretized lattice models in two and three dimensions.
Section~\ref{sec:conclusions} summarizes how \freebird serves as a versatile platform for constructing and modeling interfacial systems, and how its functionality is expected to expand to support additional models, positioning it as a focal point for future surface-related modeling and computational materials design.
This work is not intended as a comprehensive documentation of the \freebird package (which is provided separately) but rather as a pedagogical blueprint of its composition, intended to convey the underlying software design philosophy and scientific vision to future users and developers.

\section{Implementations\label{sec:imp}}

In this section, we outline the design philosophy and structure of \freebird, along with the data types it implements to enable efficient execution.
We then describe how these data types support the construction of computational models for both continuous atomistic and discretized lattice systems.
Finally, we summarize the configurational sampling methods currently available in \freebird, providing a brief introduction to technical details relevant to our demonstrations. \textblue{All results reported here were obtained using \freebird release version 0.2.1.}

\subsection{Code}

Here, we provide an overview of \freebird's structure and key functionalities.
The code is designed for readability and maintainability through a highly modular, functional architecture, and achieves performance by introducing specialized data types and enforcing type stability within functions.
We also enhance sustainability and reproducibility by following established research software engineering best practices.
This section does not attempt to detail all aspects of code usage; comprehensive documentation, including examples and tutorials, is available at \url{https://wexlergroup.github.io/FreeBird.jl/}, which should be consulted for the most up-to-date information.

\subsubsection{Overview of the Code}

The \freebird source code is organized into several modules, each containing functions related to a specific theme or topic.
These modules fall into a few broad categories, many of which define Julia \textit{abstract types} to establish the conceptual hierarchy of data types.
The first category comprises modules for system construction, where new data types are introduced:
\begin{itemize}
    \item \codeword{AbstractWalkers}: Defines the abstract concept of walkers, the central data structure in \freebird, which stores system information such as atomic or lattice configuration, number of components, and energy.
    \item \codeword{AbstractLiveSets}: Defines the abstract concept of live sets (from nested sampling terminology), which group a collection of walkers with an associated energy-evaluating function. More generally, a live set can be viewed as a data type holding both a list of walkers and an energy calculator.
    \item \codeword{AbstractPotentials}: Defines abstract types for interatomic potentials, such as the Lennard-Jones (LJ) potential.
    \item \codeword{AbstractHamiltonians}: Defines abstract types for discrete Hamiltonians, designed for efficient energy evaluation of lattice systems.
\end{itemize}
\vspace{2mm}

The next category comprises functional modules, code that operates on the data types described above to perform sampling or enumeration:
\begin{itemize}
    \item \codeword{SamplingSchemes}: Defines sampling methods, including nested sampling, Wang--Landau sampling, Monte Carlo sampling, and exact enumeration.
    \item \codeword{MonteCarloMoves}: Provides definitions of Monte Carlo moves, such as random walks and swap moves, for use within the sampling methods.
    \item \codeword{EnergyEval}: Specifies how energies are computed by combining a walker with an abstract potential or Hamiltonian.
\end{itemize}
\vspace{2mm}

The penultimate category includes a dedicated module for input/output handling:
\begin{itemize}
    \item \codeword{FreeBirdIO}: Provides functions for converting between atomic structures stored in files and the walker system defined in \freebird. It also defines \codeword{DataSavingStrategy} types, which specify how and when output data are saved during a sampling run, as well as the format of the saved data.
\end{itemize}
\vspace{2mm}

Lastly, we include a module providing auxiliary tools:
\begin{itemize}
    \item \codeword{AnalysisTools}: Offers convenience functions for post-calculation analyses. This module is optional, and the code can operate without it.
\end{itemize}
\vspace{2mm}

\begin{figure}[tb]
    \centering
    \includegraphics[width=1\linewidth]{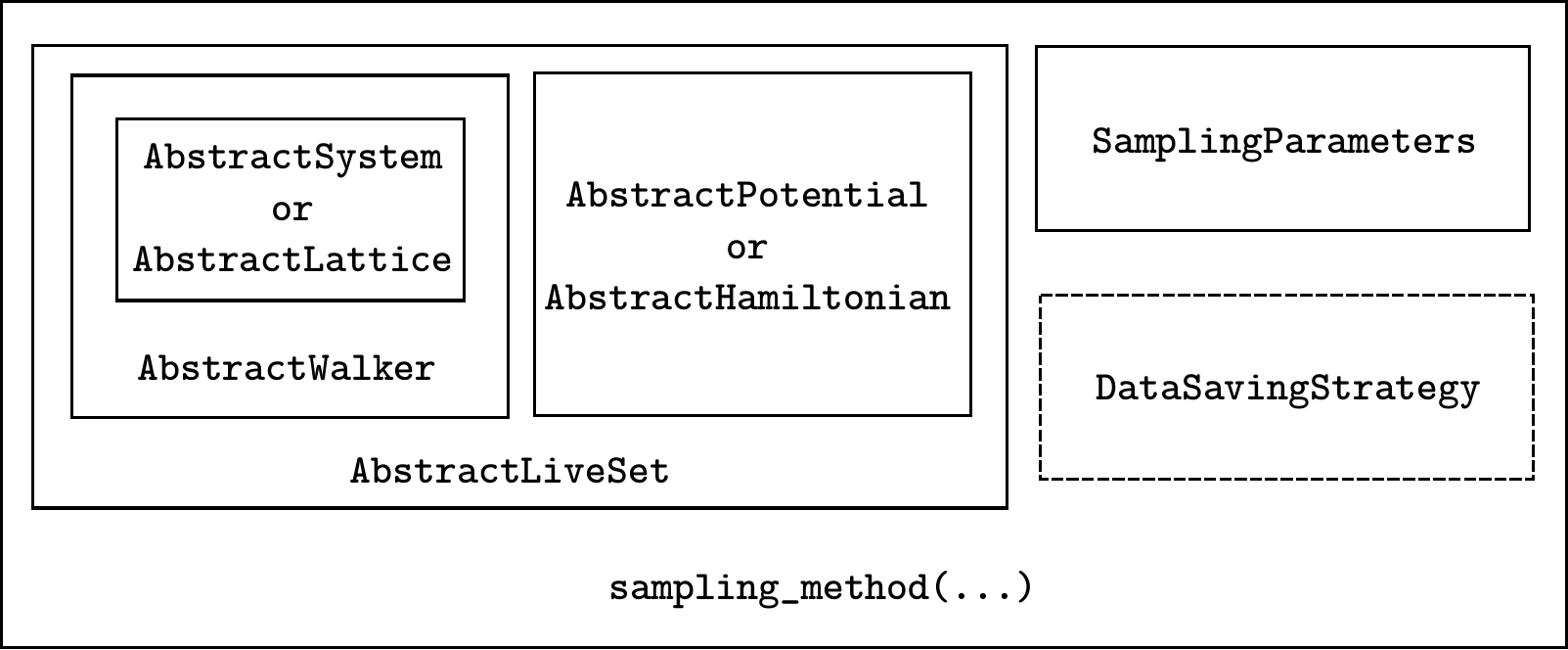}
    \caption{
\textbf{Schematic overview of the \freebird data structure.}
To perform a sampling calculation, the required settings are specified in \codewordcaption{SamplingParameters}.
Typically, either a live set or a single initial walker, containing the embedded system description, is used.
The system is represented by an \codewordcaption{AbstractSystem} or \codewordcaption{AbstractLattice}, wrapped by an \codewordcaption{AbstractWalker}.
Energies are evaluated using an \codewordcaption{AbstractPotential} or \codewordcaption{AbstractHamiltonian}, which is attached to the live set or walker.
Optional data output behavior is configured via a \codewordcaption{DataSavingStrategy}.
}
    \label{fig:concept}
\end{figure}

Figure~\ref{fig:concept} provides an overview of how the data types defined in the modules are integrated to perform a sampling calculation.
Further details about the data types themselves are presented in the following section.

\subsubsection{Data Structure and Type Systems}

As a just-in-time (JIT) compiled language, Julia can generate specialized, optimized machine code for fast runtime execution when variable data types are explicitly specified in functions.
Julia \textit{functions} can also perform different tasks depending on the data types of their inputs, a paradigm known as \textit{multiple dispatch}.

\codeword{AbstractWalker} is the core data type in \freebird, with two current subtypes: \codeword{AtomWalker{C}} and \codeword{LatticeWalker{C}}.
An \codeword{AbstractWalker} object stores both the system's structure or configuration and its energy, along with supplementary data such as iteration number and particle constraints.
Both \codeword{AtomWalker{C}} and \codeword{LatticeWalker{C}} are parameterized by \codeword{C}, the number of components in the system, enabling specialized function implementations.
For example, single-component systems are simpler, allowing dedicated, faster methods for \codeword{AtomWalker{1}} that avoid the component loops required when \codeword{C}~$>1$.

\codeword{AbstractPotential} and \codeword{AbstractHamiltonian} are abstract types for defining interactions.
For example, \codeword{LennardJonesParameterSets} is a subtype of \codeword{AbstractPotential} that stores one or more parameter sets for defining an LJ potential.
Similarly, \codeword{GenericLatticeHamiltonian} is a subtype of \codeword{AbstractHamiltonian} that specifies an on-site interaction energy and a list of $n$-th nearest-neighbor interactions.

\codeword{AbstractLiveSet} is a data type that holds a list of \codeword{AbstractWalker}s together with an \codeword{AbstractPotential} or \codeword{AbstractHamiltonian}.
When these elements are combined, the energies of the \codeword{AbstractWalker}s are automatically calculated and updated.
This type is central to performing nested sampling and also serves as a convenient structure for organizing inputs and outputs from other sampling methods.

Type stability is critical for the performance of Julia code.
It means that a function's output type depends on the \textit{types} of its inputs rather than their \textit{values}.
This property enables the Julia compiler to determine the output type at compile time, allowing efficient memory allocation at runtime.
If the return type cannot be inferred, the machine code must accommodate all possible outcomes, which significantly slows execution.
Nearly all functions in \freebird are designed to be type stable, except for those in auxiliary modules that are not used during calculations.

\subsubsection{Research Software Development Best Practices}

Continuous integration (CI) is a central component of \freebird's development process.
Ensuring research software is maintainable and its development sustainable requires deliberate investment in and enforcement of good practices.
In this section, we outline the software development practices adopted in \freebird and describe how they are automated.

We use GitHub for version control, enforcing strict branch policies.
New features are developed in a \texttt{feature/*} branch and merged into \texttt{dev} only after passing tests and receiving reviewer approval via a pull request.
The \texttt{dev} branch is subsequently merged into \texttt{main}, typically in conjunction with a new release.
\texttt{hotfix/*} branches are reserved for bug fixes and are merged promptly into both \texttt{dev} and \texttt{main}.
\texttt{docs/*} branches are used exclusively for documentation-related changes, including updates to the dedicated \texttt{docs} directory and function docstrings in the source code.

\codeword{Documenter.jl} \cite{documenter_jl} is used to automatically generate and deploy \freebird's documentation to GitHub Pages.
It collects and renders method docstrings for inclusion in the documentation, ensuring that function descriptions remain synchronized with the source code.
All exported methods are required to have docstrings describing their functionality, input arguments, and return values (if any).
Documentation for any function can also be accessed directly from the Julia REPL.
\codeword{Documenter.jl} builds separate pages for stable, development, and release versions of the code, and generates a preview when a pull request is opened.
The \freebird documentation also includes tutorials and examples to support new users and developers.

Unit testing is a critical component of CI.
We use Julia's built-in testing framework to create test suites for each module of \freebird and employ \href{https://coveralls.io/}{\textit{Coveralls}} to track and report code coverage.
Maintaining equal or higher coverage is a requirement for pull request approval.

Other automated workflows include the \href{https://github.com/JuliaRegistries/TagBot}{TagBot}, which automatically creates tags, releases, and changelogs when a new version of the code is registered in the Julia General registry.
\href{https://github.com/JuliaRegistries/CompatHelper.jl}{CompatHelper} periodically updates \freebird's dependencies, generating an automated pull request whenever a new version of a dependency package becomes available.

All CI components are automated using GitHub Actions.
The workflow files, located in the \texttt{.github/workflows} directory of the \freebird repository, can be readily incorporated into any Julia-based project.
To ensure readability and consistency, we follow the \href{https://github.com/JuliaDiff/BlueStyle}{BlueStyle} coding conventions when developing \freebird.
Overall, Julia offers a mature and comprehensive CI ecosystem, well integrated with GitHub and other platforms, that supports the sustainable development of \freebird.

\subsubsection{Parallelization}

Another aspect of \freebird's flexibility is its ability to perform fast, lightweight computations with relatively inexpensive energy evaluators on limited computing resources, while also scaling to high-performance computing (HPC) platforms for intensive, large-scale campaigns.
Currently, \freebird uses two main forms of parallelism (\codeword{Threads} for multithreading and \codeword{Distributed} for multiprocessing), both native to the Julia standard library, to accelerate computations across available hardware.
\codeword{Threads} are used for embarrassingly parallel loops, while \codeword{Distributed} is applied to sampling steps that can be executed concurrently (primarily in nested sampling; see Section~\ref{sec:algo-ns}) by spawning multiple \codeword{worker}s, typically one per CPU core.
Both approaches can be combined by assigning multiple threads to each \codeword{worker} to further increase CPU utilization, as most \freebird functions are threaded.

Julia supports parallelization via the Message Passing Interface (MPI) protocol through external packages such as \codeword{MPI.jl}. \cite{mpi_jl}
GPU computing is also enabled by hardware-specific packages, including \codeword{CUDA.jl} \cite{besard2018juliagpu, besard2019prototyping, cuda_jl} for NVIDIA, \codeword{AMDGPU.jl} \cite{amdgpu_jl} for AMD, \codeword{oneAPI.jl} \cite{oneapi_jl} for Intel, and \codeword{Metal.jl} \cite{metal_jl} for Apple Silicon GPUs.
Additionally, vendor-neutral GPU programming models such as \codeword{OpenCL.jl} \cite{opencl_jl} and \codeword{KernelAbstractions.jl} \cite{Churavy_KernelAbstractions_jl} allow Julia code to run across multiple platforms from different hardware vendors.

\subsection{Models}

In \freebird, physical models are classified into two categories: (1) \textbf{atomistic systems}, composed of particles with positions that vary continuously in real space (typically $\mathbb{R}^{3}$), where the total energy is defined by a Hamiltonian derived from an interatomic potential; and (2) \textbf{lattice systems}, in which particles occupy a discrete set of lattice sites and the energy is defined by a lattice Hamiltonian.
Both categories can accommodate any number of chemical species and arbitrary cell shapes, with or without periodic boundaries, either in all spatial dimensions or in a user-selected subset of dimensions.
\freebird also provides numerous convenience functions to construct these systems (see examples later in this section).

{%\color{blue}
Although the present work highlights applications to monoatomic solids, forthcoming releases will extend \freebird to molecular systems by integrating ASE calculators \cite{herbst2025b} for classical \cite{jorgensen1988opls, cornell1995second, mackerell1998all} and machine-learning \cite{zhang2025ani, anstine2025aimnet2, wood2025family} force fields and coupling with \codeword{Molly.jl} \cite{Greener2024} to enable molecular dynamics–based sampling of bonds, angles, and dihedrals.

We provide code examples to illustrate the typical Julia syntax used in \freebird, but these listings are entirely optional. Skipping them will not affect the overall comprehensibility of the work.
}

\begin{figure}[t!]
\begin{singlespace} 
\begin{minipage}{0.475\textwidth}
\begin{listing}[H]
\begin{mdframed}
\begin{minted}[linenos,breaklines,fontsize=\scriptsize]{julia}
# Load FreeBird
julia> using FreeBird

# Generate 3 random configurations of 5 hydrogen atoms in a cubic box with 10.0 Å³ volume per atom
julia> hydrogens = generate_initial_configs(3, 10.0, 5)
3-element Vector{AtomsBase.FastSystem{...}}:
 FastSystem(H₅, periodicity = FFF)
 FastSystem(H₅, periodicity = FFF)
 FastSystem(H₅, periodicity = FFF)

# Draw an ASCII diagram of the first configuration
julia> view_structure(hydrogens[1])
   .--------.
  /|        |
 * |  H     |
 H |        |
 | |     H  |
 | .--H----H.
 |/        /
 *--------*

# Wrap the first configuration in a two-component walker; split the 5 atoms into components of 2 and 3 atoms, and freeze the first component
julia> wk = AtomWalker{2}(hydrogens[1]; list_num_par=[2,3], frozen=[true,false])
AtomWalker{2}(
    configuration      : FastSystem(H₅, periodicity = FFF)
    energy             : 0.0 eV
    iter               : 0
    list_num_par       : [2, 3]
    frozen             : Bool[1, 0]
    energy_frozen_part : 0.0 eV)

# Use broadcasting to wrap all configurations in `hydrogens` into single-component walkers, with all atoms free
julia> wks = AtomWalker.(hydrogens)
3-element Vector{AtomWalker{1}}:
 AtomWalker{1}(
    configuration      : FastSystem(H₅, periodicity = FFF)
    energy             : 0.0 eV
    iter               : 0
    list_num_par       : [5]
    frozen             : Bool[0]
    energy_frozen_part : 0.0 eV)
...
\end{minted}
\end{mdframed}
\end{listing}
\end{minipage}
\end{singlespace}
\caption{\label{code:walkers}
Julia code demonstrating the generation of three \codewordcaption{AtomWalker} objects for systems containing five hydrogen atoms in a cubic box, including visualization, multi-component splitting with partial freezing, and batch wrapping via broadcasting.
Some output is truncated.
}
\end{figure}

\subsubsection{Atomistic Systems with Interatomic Potentials}

We use the term \textit{atomistic} to describe a system composed of individual particles whose positions vary continuously in real space, confined within a simulation cell that may employ periodic boundary conditions.
Configuration space is explored using \textit{single-particle displacement moves}, trial translations with directions chosen uniformly at random and magnitudes drawn from a prescribed distribution.
The sequence of accepted moves forms a random walk through configuration space, which is the default sampling scheme in \freebird (among others).
Figure~\ref{code:walkers} illustrates how to generate a set of random atomistic configurations and store them in an \codeword{AtomWalker{C}} data type.

\paragraph{Construction of Atomistic Configurations.}

As noted earlier, in \codeword{AtomWalker{C}}, \codeword{C} is a positive integer specifying the number of components (e.g., distinct chemical species or rigid subgroups) represented by the walker.
In Figure~\ref{code:walkers}, line~23 demonstrates how a system of five particles in a cubic box can be divided into two components, with the first component frozen.
Frozen particles are excluded from subsequent single-particle displacement moves; their energy is computed once and stored, while interactions with mobile particles are still calculated on the fly, thereby reducing overall computational cost.

In an atomistic system, components are defined as groups of particles sharing a common characteristic (for example, distinct chemical species, rigid fragments, or subsets subject to identical move or interaction parameters).
Many functions in \freebird operate natively on such multi-component systems; for instance, the total potential energy is computed as the sum of inter- and intra-component contributions.
This design provides flexibility for modeling diverse systems, including mixtures, host--guest complexes, and interfaces.

\paragraph{Definition of Atomistic Potentials.}

{%\color{blue}
Particles in an atomistic system interact through interatomic potentials, implemented as subtypes of \codeword{AbstractPotential}, currently classified as single- or multi-component and as either pairwise or many-body.
For example, in Figure~\ref{code:liveset}, line~2, the \codeword{LJParameters} (a single-component, pairwise potential) is constructed using keyword arguments.}
In this example, specifying a potential cutoff at $4\sigma$ automatically shifts the potential energies so they approach 0~eV at the cutoff.
This shift can be disabled by supplying the additional argument \codeword{shift=false}.
{%\color{blue}
For multi-component systems, \codeword{CompositeParameterSets} can be used to define interactions.
This data structure, a subtype of \codeword{MultiComponentPotential}, stores a matrix of \codeword{LJParameters}.
}

Lastly, Figure~\ref{code:liveset} illustrates how to construct a live set by combining a set of walkers with an LJ potential.
\codeword{LJAtomWalkers}, a subtype of \codeword{AtomWalkers} (itself a subtype of \codeword{AbstractLiveSet}), contains \codeword{AtomWalker} configurations of particles interacting via LJ potentials.
By default, when walkers and a potential are passed to \codeword{LJAtomWalkers}, their energies are automatically computed and updated.
For example, the walkers in \codeword{wks} in line~33 of Figure~\ref{code:walkers} initially have no assigned energies; invoking \codeword{LJAtomWalkers(wks, lj)} in line~6 of Figure~\ref{code:liveset} evaluates those energies using the potential \codeword{lj} defined in line~2.
This design allows any set of walkers to be paired with any potential while keeping energies current.
Supplying the keyword argument \codeword{assign_energy=false} disables this feature, leaving energies unchanged.
Many functions in \freebird follow this pattern, where expected default behavior can be overridden through additional user-provided arguments.
This approach supports robust code execution and reliable user interaction while enabling developers to explore unconventional settings.

\begin{figure}[t!]
\begin{singlespace} 
\begin{minipage}{0.475\textwidth}
\begin{listing}[H]
\begin{mdframed}
\begin{minted}[linenos,breaklines,fontsize=\scriptsize]{julia}
# Construct a Lennard-Jones potential
julia> lj = LJParameters(epsilon=0.1, sigma=2.5, cutoff=4.0)
LJParameters(0.1 eV, 2.5 Å, 4.0, -9.763240814208984e-5 eV)

# Combine the walkers and Lennard-Jones potential into a live set
julia> ls = LJAtomWalkers(wks, lj)
LJAtomWalkers(AtomWalker{1}, LJParameters):
[1] AtomWalker{1}(
    configuration      : FastSystem(H₅, periodicity = FFF)
    energy             : 496.83218514161376 eV
    iter               : 0
    list_num_par       : [5]
    frozen             : Bool[0]
    energy_frozen_part : 0.0 eV)

[2] AtomWalker{1}(
    configuration      : FastSystem(H₅, periodicity = FFF)
    energy             : 75.15053441600284 eV
    iter               : 0
    list_num_par       : [5]
    frozen             : Bool[0]
    energy_frozen_part : 0.0 eV)

[3] AtomWalker{1}(
    configuration      : FastSystem(H₅, periodicity = FFF)
    energy             : 5016.408030136223 eV
    iter               : 0
    list_num_par       : [5]
    frozen             : Bool[0]
    energy_frozen_part : 0.0 eV)

LJParameters(0.1 eV, 2.5 Å, 4.0, -9.763240814208984e-5 eV)
\end{minted}
\end{mdframed}
\end{listing}
\end{minipage}
\end{singlespace}
\caption{\label{code:liveset}
Julia code for constructing a Lennard-Jones potential and combining it with the walkers from Figure~\ref{code:walkers} to create a live set (\codewordcaption{LJAtomWalkers}).
Energies correspond to the initial configurations; output is shown in full.
}
\end{figure}

\subsubsection{Lattice Systems with Discrete Hamiltonians\label{sec:lattice-model}}

We implement lattice models with two primary objectives: (1) to serve as coarse-grained alternatives to continuous atomistic models, and (2) to enable the study of systems whose exact partition function can be computed under certain simplifying conditions (e.g., small lattices or Ising-like systems), thereby providing accuracy benchmarks for the sampling methods.

\begin{figure}[b!]
\begin{singlespace} 
\begin{minipage}{0.475\textwidth}
\begin{listing}[H]
\begin{mdframed}
\begin{minted}[linenos,breaklines,fontsize=\scriptsize]{julia}
# Construct a single-component lattice with 4 particles occupying the first 4 lattice sites using SLattice (alias for MLattice with one component)
julia> sl = SLattice{SquareLattice}(components=[[1,2,3,4]])
SLattice{SquareLattice}
    lattice_vectors      : [1.0 0.0 0.0; 0.0 1.0 0.0; 0.0 0.0 1.0]
    positions            : 16 grid points
    supercell_dimensions : (4, 4, 1)
    basis                : [(0.0, 0.0, 0.0)]
    periodicity          : (true, true, false)
    cutoff radii         : 2 nearest neighbors cutoffs [1.1, 1.5]
    occupations          :
      ● ● ● ●
      ○ ○ ○ ○
      ○ ○ ○ ○
      ○ ○ ○ ○
    adsorptions          : full adsorption

# Construct a two-component lattice (MLattice requires the number of components as a type parameter). Split the 4 particles into two components: [1,2] and [3,4]
julia> ml = MLattice{2,TriangularLattice}(components=[[1,2], [3,4]])
MLattice{2, TriangularLattice}
    lattice_vectors      : [1.0 0.0 0.0; 0.0 1.7320508075688772 0.0; 0.0 0.0 1.0]
    positions            : 16 grid points
    supercell_dimensions : (4, 2, 1)
    basis                : [(0.0, 0.0, 0.0), (0.5, 0.8660254037844386, 0.0)]
    periodicity          : (true, true, false)
    cutoff radii         : 2 nearest neighbors cutoffs [1.1, 1.5]
    occupations          :
      ➊ ➋ ○ ○
       ➊ ➋ ○ ○
      ○ ○ ○ ○
       ○ ○ ○ ○
    adsorptions          : full adsorption
\end{minted}
\end{mdframed}
\end{listing}
\end{minipage}
\end{singlespace}
\caption{\label{code:sl}
Julia code for constructing a single-component square lattice (\codewordcaption{sl}) and a two-component triangular lattice (\codewordcaption{ml}) with four occupied sites.
The two-component system assigns particles [1,2] to the first component and [3,4] to the second.
}
\end{figure}

\paragraph{Construction of Lattice Configurations.}

Lattice configurations are represented by the built-in, parameterized data type \codeword{MLattice{C,G}}, where \codeword{C} specifies the number of components (in the chemical thermodynamic sense) and \codeword{G} denotes the lattice geometry.
\codeword{MLattice{C,G}} is a subtype of \codeword{AbstractLattice}.
Figure~\ref{code:sl} provides two examples of constructing two-dimensional lattice systems in \freebird: (1) a single-component square lattice (\codeword{C=1} by default, \codeword{G=SquareLattice}) created using \codeword{SLattice{G}}, an alias for \codeword{MLattice{1,G}}, and (2) a two-component triangular lattice (\codeword{C=2}, \codeword{G=TriangularLattice}).
In both cases, only the keyword argument \codeword{components} is passed to the constructor, with all other fields (lattice vectors, basis, and supercell dimensions) set to their default values but modifiable by the user.
To model adsorption processes, a subset of lattice grid points can be designated as adsorption sites.
Particles occupying these sites experience an on-site energy in addition to their neighboring interaction energies.

\paragraph{Definition of Lattice Hamiltonians.}

The energy of a lattice can be efficiently evaluated using a discrete Hamiltonian, which specifies only the on-site energy and as many $n$-th nearest-neighbor interactions as needed.
Because a lattice is defined on a discretized grid rather than by continuous distances, a lattice gas model is considerably more efficient than an interatomic potential, where energy depends explicitly on distance.
For example, Figure~\ref{code:ham} shows the construction of a generic lattice Hamiltonian, specifying the magnitudes of the on-site, first-nearest, and second-nearest neighbor interactions, along with the units of energy.
The lattice Hamiltonian is given by
\begin{equation}
    H = N \varepsilon_0 + \sum_{\mathrm{nn}} \varepsilon_{\mathrm{nn}} + \sum_{\mathrm{nnn}} \varepsilon_{\mathrm{nnn}} + \cdots,
\end{equation}
where $N$ is the number of occupied sites, $\varepsilon_0$ is the energy of an occupied site, $\varepsilon_{\mathrm{nn}}$ is the nearest-neighbor interaction energy, and $\varepsilon_{\mathrm{nnn}}$ is the next-nearest-neighbor interaction energy.
We pass these interaction values into a \codeword{GenericLatticeHamiltonian{N,U}}, as shown in Figure~\ref{code:ham}, where \codeword{N} is the number of $n$-th nearest-neighbor interaction terms included (two in this example) and \codeword{U} specifies the energy units.
The Hamiltonian is independent of specific lattice geometry and can therefore be applied to any lattice type.
For multi-component systems, a matrix of \codeword{GenericLatticeHamiltonian} objects can be defined for inter- and intra-component interactions using \codeword{MLatticeHamiltonian{C,N,U}}, where \codeword{C} is the number of components.

\begin{figure}[t!]
\begin{singlespace} 
\begin{minipage}{0.475\textwidth}
\begin{listing}[H]
\begin{mdframed}
\begin{minted}[linenos,breaklines,fontsize=\scriptsize]{julia}
# Define the Hamiltonian:
# on-site energy = -0.04 eV
# nearest-neighbor energy = -0.01 eV
# next-nearest-neighbor energy = -0.0025 eV
julia> h = GenericLatticeHamiltonian(-0.04, [-0.01, -0.0025], u"eV")
GenericLatticeHamiltonian{2,Quantity{Float64, L² MT⁻², Unitful.FreeUnits{(eV,), L² MT⁻², nothing}}}:
    on_site_interaction:      -0.04 eV
    nth_neighbor_interactions: [-0.01, -0.0025] eV
\end{minted}
\end{mdframed}
\end{listing}
\end{minipage}
\end{singlespace}
\caption{\label{code:ham}
Julia code for constructing a \codewordcaption{GenericLatticeHamiltonian} with specified on-site, nearest-neighbor, and next-nearest-neighbor interaction energies.
}
\end{figure}

\subsection{Methods}

In this section, we provide brief introductions to several sampling algorithms currently implemented in \freebird, including Metropolis sampling, Wang--Landau sampling, nested sampling, and exact enumeration for finite lattices.
Serving as a toolbox, \freebird is continually expanding its range of sampling approaches; here, we describe only those currently available.
Since these algorithms are regularly optimized and updated, we focus here on outlining their underlying fundamental principles rather than implementation-specific details.

\subsubsection{Metropolis Sampling}

The Monte Carlo (MC) sampling in this study follows the Metropolis algorithm. \cite{metropolis_equation_1953}
This method samples from the canonical ensemble at fixed temperature by generating a random walk through configuration space.
The probability of transitioning from state 1 to state 2 is given by
\begin{equation} \label{equation-for-mc-probability}
    P(1\rightarrow2)=\min\left[1,e^{-\beta(E_2-E_1)}\right],
\end{equation}
where $E_1$ and $E_2$ are the energies of states 1 and 2, respectively; $k_{\mathrm{B}}$ is the Boltzmann constant; $T$ is the temperature; and $\beta=1/(k_{\mathrm{B}}T)$ is the inverse temperature.
The algorithm proceeds as follows:
\vspace{2mm}
\begin{enumerate}
    \item
Initialize the configuration.
    \item \label{step-for-mc-move-proposal}
Select a particle at random and propose a move:
    \begin{itemize}
        \item
\textit{Atomistic}: Displace by a random vector.
        \item
\textit{Lattice}: Swap with a randomly chosen empty site.
    \end{itemize}
    \item
Compute the energy difference $\Delta E=E_2-E_1$.
    \item \label{step-for-mc-move-acceptance}
Accept the move with probability $P(1\rightarrow2)$, as given in eq~\ref{equation-for-mc-probability}.
    \item
Repeat steps \ref{step-for-mc-move-proposal}--\ref{step-for-mc-move-acceptance} for many iterations.
\end{enumerate}
\vspace{2mm}
Thermodynamic properties, such as the constant-volume heat capacity $C_V(T)$, can be obtained by analyzing the resulting ensemble.
In \freebird, the only required inputs are the number of iterations and, for atomistic systems, the step size.
The package also supports temperature sweeps, where the final configuration at one temperature is used as the initial configuration for the next, facilitating efficient re-equilibration.

\subsubsection{Wang--Landau Sampling}

The Wang--Landau (WL) algorithm is an MC method that achieves uniform sampling across a system's energy range by iteratively refining an estimate of the density of states, $g(E)$. \cite{wang_efficient_2001}
This adaptive strategy enables thorough exploration of configuration space, including rugged or frustrated energy landscapes where conventional Metropolis algorithms may become trapped in metastable states.
The algorithm proceeds as follows:
\vspace{2mm}
\begin{enumerate}
    \item
Initialize the configuration, $g(E)$, energy histogram $H(E)$, and modification factor $f$.
    \item \label{step-for-wl-move-proposal}
Select a particle at random and propose a Metropolis-style move.
    \item
Compute the ratio $\eta=g(E_1)/g(E_2)$.
    \item
Accept the move with $P(1\rightarrow2)=\min\left[1,\eta\right]$.
    \item
Update $g$ and $H$ at the visited energy $E_{\mathrm{v}}$:
    \begin{itemize}
        \item
$g(E_{\mathrm{v}}) \leftarrow g(E_{\mathrm{v}}) \times f$
        \item
$H(E_{\mathrm{v}}) \leftarrow H(E_{\mathrm{v}})+1$
    \end{itemize}
    \item \label{step-for-wl-flatness-check}
If $H(E)$ is sufficiently flat, reset $H(E)$ and reduce $f$.
    \item
Repeat steps \ref{step-for-wl-move-proposal}--\ref{step-for-wl-flatness-check} until $f$ meets the convergence criterion.
\end{enumerate}
\vspace{2mm}
This method provides direct estimates of $g(E)$, from which thermodynamic quantities such as entropy and free energy can be derived.
In \freebird, additional required inputs beyond those for Metropolis sampling include the energy range and resolution, as well as convergence criteria for $H(E)$ flatness and $f$ reduction.

\subsubsection{Nested Sampling\label{sec:algo-ns}}

The nested sampling (NS) algorithm is an MC method in which the accessible phase space contracts stochastically, producing approximately uniform steps in $\ln\Gamma$, the logarithm of the remaining phase-space volume. \cite{skilling_nested_2004, skilling_nested_2006}
This transformation converts the high-dimensional canonical partition function integral into a tractable one-dimensional sum. \cite{partay_efficient_2010, partay_nested_2021, ashton_nested_2022}
The algorithm proceeds as follows:
\vspace{2mm}
\begin{enumerate}
    \item
Initialize the \textit{live set} of $K$ \textit{walkers}, each representing a configuration sampled uniformly from the prior distribution over allowed configurations.
    \item \label{step-for-ns-walker-selection}
Select the $C$ highest-energy walkers.
    \item
Update the energy limit to $E_{\mathrm{lim}} \leftarrow E_{\mathrm{highest}}$.
    \item
Compute the expected cumulative configuration-space contraction after $i$ iterations as $\langle\Gamma_i\rangle=\left[K/(K+C)\right]^i$.
    \item \label{step-for-ns-walker-replacement}
Replace the $C$ selected walkers with new configurations drawn uniformly from the region where $E_{\mathrm{new}} \le E_{\mathrm{lim}}$.
    \item
Repeat steps~\ref{step-for-ns-walker-selection}--\ref{step-for-ns-walker-replacement} for many iterations.
\end{enumerate}
\vspace{2mm}
The procedure continues until sufficiently low-energy regions of configuration space have been sampled.
From the sequence of replaced walkers and their associated energies, the canonical partition function is estimated as
\begin{equation} \label{equation-for-ns-Z}
    Z(\beta)=\sum_i w_i e^{-\beta E_i},
\end{equation}
where $w_i=\Gamma_{i-1}-\Gamma_i$ is the weight for the $i$-th iteration.
Because the sampling steps are temperature-independent, any value of $\beta$ can be substituted into eq.~\eqref{equation-for-ns-Z} during post-processing, allowing thermodynamic properties to be computed over a broad temperature range from a single simulation.
In \freebird, additional required inputs beyond those for Metropolis sampling include $K$ walkers, $C$, and the total number of NS iterations.

\subsubsection{Exact Enumeration for Lattice Systems}

Exact enumeration (also referred to as \textit{complete}, \textit{direct}, or \textit{exhaustive} enumeration) is the process of explicitly generating all possible configurations (i.e., all arrangements of particles and their species) on a finite, discretized system such as a lattice with a fixed number of sites and particles.
Each unique microstate is identified and its energy evaluated to compute the exact partition function of the system.
The method is straightforward to implement, typically by generating all unique permutations of an integer array, for example:
\begin{align*}
    \underbrace{[1,1,1,2,2,3,3,\dots,0,0,0]}_{\text{number\ of\ lattice\ sites}},
\end{align*}
where nonzero entries label (chemical) species and zeros denote unoccupied sites.
Because it enumerates the full microstate space, this approach serves as a reference standard for benchmarking approximate sampling methods.
Its applicability, however, is limited to small lattice systems, as the number of configurations grows exponentially with the total number of particles and sites.

\section{Demonstrations\label{sec:demo}}

\begin{figure*}[tb]
\centering
\begin{mdframed}
\begin{singlespace} 
\begin{minipage}{1\textwidth}
\begin{minted}[linenos,breaklines,fontsize=\scriptsize]{julia}
# Generate an initial set of 120 walkers and an LJ potential
walkers = AtomWalker.(generate_initial_configs(120, 562.5, 6))
lj = LJParameters(epsilon=0.1, sigma=2.5, cutoff=4.0)

# Nested sampling
ns_energies, ns_ls, ns_params = nested_sampling(LJAtomWalkers(walkers, lj), NestedSamplingParameters(mc_steps=200, step_size=0.1), 30_000, MCRandomWalkClone(), SaveEveryN(n_traj=10))

# Metropolis Monte Carlo using a temperature grid from 1000 K down to 50 K, in 50 K steps
mc_energies, mc_ls, mc_cvs, acceptance_rates = monte_carlo_sampling(walkers[1], lj, MetropolisMCParameters(collect(1000.0:-50:50), equilibrium_steps=100_000, sampling_steps=100_000, step_size=0.1))

# Wang-Landau sampling
wl_energies, wl_ls, wl_params, S, H = wang_landau(walkers[1], lj, WangLandauParameters(num_steps=10_000, energy_min=-1.26, energy_max=0.0, num_energy_bins=1_000, step_size=1.0, f_min=1.00001))
\end{minted}
\end{minipage}
\end{singlespace}
\centering
\includegraphics[width=0.8\linewidth]{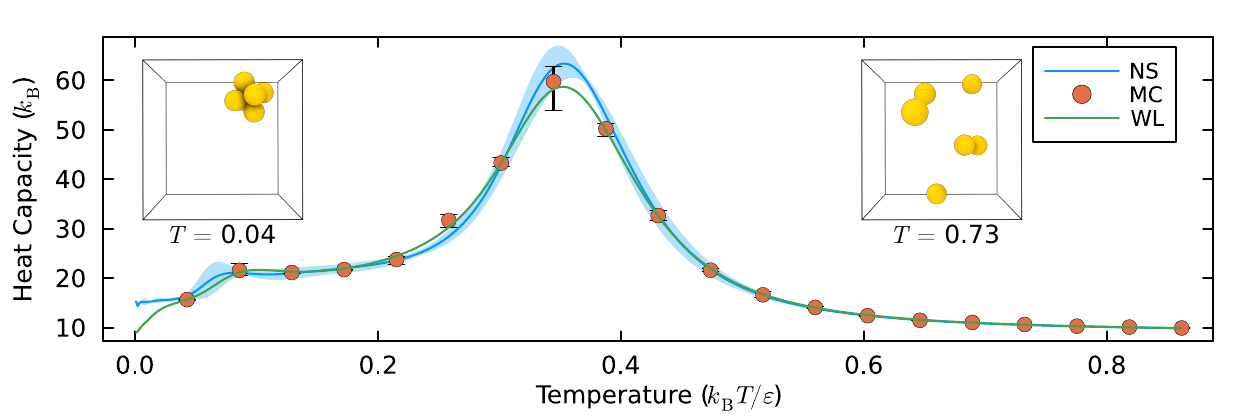}
\end{mdframed}
\caption{
Julia code and results for the constant-volume heat capacity of the LJ$_6$ cluster computed using nested sampling (NS), Metropolis Monte Carlo (MC), and Wang--Landau (WL) sampling.
Each method was repeated three times; NS and WL are shown as lines with shaded ranges (min--max), and MC as markers with error bars (min--max).
Insets show representative structures from high ($0.73~k_{\mathrm{B}}T/\varepsilon$) to low ($0.04~k_{\mathrm{B}}T/\varepsilon$) temperature, illustrating the gas-to-cluster transition.
Structures are taken from NS and verified with MC at corresponding temperatures.
}
\label{fig:LJ6}
\end{figure*}

In this section, we demonstrate how \freebird can be used to identify the number, temperature, and nature of phase transitions across diverse systems.
These span single- and multi-component compositions, spatial resolutions from continuous atomistic models to discretized lattice models, and dimensionalities in both two and three dimensions.
We also benchmark three sampling methods (NS, MC, and WL) to illustrate \freebird's versatility and to validate its predictions.
The complete set of simulation parameter values is provided in the Julia code associated with each demonstration in Figures~\ref{fig:LJ6}--\ref{fig:lattice}.
For clarity and reproducibility, we focus on relatively simple model systems.
In each case, we compare constant-volume heat capacity (\CV) curves as functions of temperature, identifying phase transitions from the peak positions, which are determined using the \codeword{findmaxima()} function in the Julia package \codeword{Peaks.jl}.
For MC simulations, because runs were performed at discrete temperatures that may not coincide exactly with a transition temperature, the reported MC transition temperature corresponds to the simulated point nearest each identified phase transition.
Each sampling method is executed three times to estimate the uncertainty in the resulting \CV profiles.
\textblue{The reported values are the means of three runs, with the minimum and maximum at each temperature providing the lower and upper bounds of the uncertainty (hereafter referred to as min--max).}
Although these systems contain only a small number of particles, they provide minimal yet chemically and physically meaningful representations.

\subsection{Atomistic Systems}

To illustrate \freebird's applicability to models with continuously varying particle positions, we examine three representative atomistic systems:
(1) a six-particle LJ cluster (LJ$_6$),
(2) a 13-particle binary LJ cluster (A$_1$B$_{12}$), and
(3) an LJ(111) surface slab with a frozen substrate and mobile adsorbates.

\subsubsection{\texorpdfstring{LJ$_6$ Cluster}{LJ₆ Cluster}\label{demo-lj6}}

We begin with the LJ$_6$ cluster, with benchmark results available from NS. \cite{partay_efficient_2010}
We use the shifted, truncated LJ potential parameters from ref.\ \citenum{yang_surface_2024} ($\epsilon=0.1$~eV, $\sigma=2.5$~\AA, cutoff $r_c=4\sigma$).
The system consists of six free LJ particles in a non-periodic cubic box with an edge length of 15~\AA, corresponding to a number density of $2.78\times10^{-2}~\sigma^{-3}$.
We performed NS simulations with 120 walkers and a constant walker-removal rate of $C=1$ (Section~\ref{sec:algo-ns}), using the \codeword{MCRandomWalkClone()} algorithm to generate new configurations by cloning an existing walker and subsequently decorrelating it via random walks.
For MC simulations, the temperature was decreased in 50~K increments from 1000~K to 50~K.
For WL sampling, we employed 1000 bins spanning the energy range $\left[-1.26,0\right]$~eV (slightly above the ground-state energy of -1.27~eV), applied an 80~\% flatness criterion to the energy histogram, and used an $f$-schedule of $f \leftarrow \sqrt{f}$.

\begin{figure*}[tb]
\centering
\begin{mdframed}
\begin{singlespace} 
\begin{minipage}{1\textwidth}
\begin{minted}[linenos,breaklines,fontsize=\scriptsize]{julia}
# Generate an initial set of 960 walkers with two components
walkers = AtomWalker.(generate_initial_configs(960, 281.25, [1,12]; particle_types=[:H,:He]))

# Define the LJ interactions and construct a CompositeParameterSets object
lj11 = LJParameters(epsilon=0.1, sigma=2.5, cutoff=4.0)
lj22 = LJParameters(epsilon=0.05, sigma=2.5, cutoff=4.0)
lj12 = LJParameters(epsilon=sqrt(0.1*0.05), sigma=2.5, cutoff=4.0)
lj = CompositeParameterSets(2, [lj11, lj12, lj22])

# Nested sampling
ns_energies, ns_ls, ns_params = nested_sampling(LJAtomWalkers(walkers, lj), NestedSamplingParameters(mc_steps=400, step_size=0.1), 500_000, MCRandomWalkClone(), SaveEveryN(n_traj=10))

# Metropolis Monte Carlo using a temperature grid from 400 K down to 100 K, in 25 K steps
mc_energies, mc_ls, mc_cvs, acceptance_rates = monte_carlo_sampling(walkers[1], lj, MetropolisMCParameters(collect(400.0:-25:100), equilibrium_steps=10_000_000, sampling_steps=5_000_000, step_size=1.0))

# Wang-Landau sampling
wl_energies, wl_ls, wl_params, S, H = wang_landau(walkers[1], lj, WangLandauParameters(num_steps=100_000, energy_min=-2.44, energy_max=0.0, num_energy_bins=1_000, step_size=1.0, f_min=1.00001))
\end{minted}
\end{minipage}
\end{singlespace}
\centering
\includegraphics[width=0.8\linewidth]{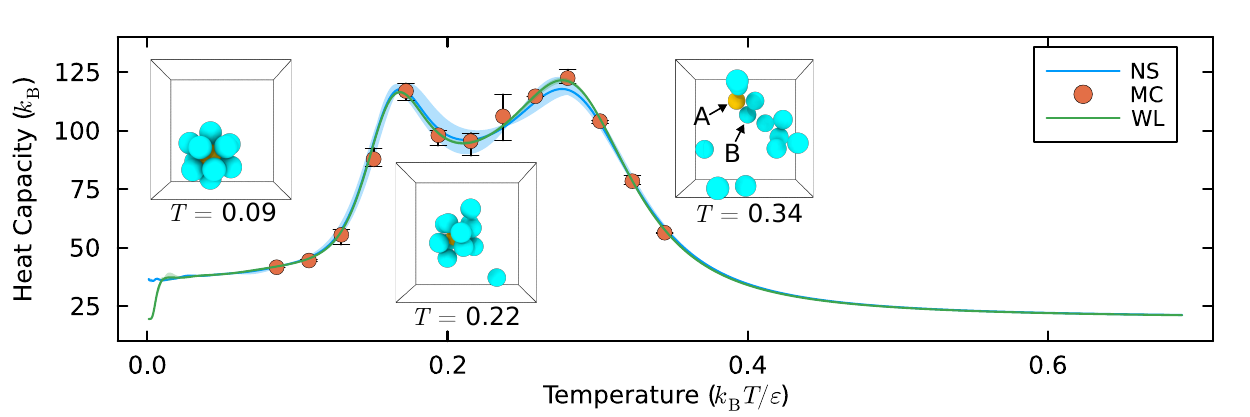}
\end{mdframed}
\caption{
Julia code and results for the constant-volume heat capacity of the LJ A$_1$B$_{12}$ cluster computed using nested sampling (NS), Metropolis Monte Carlo (MC), and Wang--Landau (WL) sampling.
Each method was repeated three times; NS and WL are shown as lines with shaded ranges (min--max), and MC as markers with error bars (min--max).
Insets show the structural change from high to low temperature, where the B-type particles condense around the single A-type particle to form a cluster.
Structures are taken from NS and corroborated by MC at corresponding temperatures.
}
\label{fig:LJ_A1B12}
\end{figure*}

Previous studies on a lower-density LJ$_6$ system ($2.31\times10^{-3}~\sigma^{-3}$) reported two phase transitions:
a gas--solid transition at $0.31~k_{\mathrm{B}}T/\varepsilon$ and
a lower-temperature solid--solid transition at $0.10~k_{\mathrm{B}}T/\varepsilon$, with the latter favoring an octahedral ground state. \cite{partay_efficient_2010}
Figure~\ref{fig:LJ6} shows that all three methods recover these transitions, with consistent transition temperatures in units of $k_{\mathrm{B}}T/\varepsilon$:
0.353 (NS), 0.345 (MC), and 0.352 (WL) for the gas--solid transition; and
0.082 (NS), 0.086 (MC), and 0.101 (WL) for the solid--solid transition. \textblue{This low-temperature transition has also been described in previous studies of small Lennard-Jones clusters as a melting from a solid to a liquid-like state, with reported transition temperatures in good agreement.} \cite{partay_efficient_2010, wales_atomic_2017, maillard_nested_2025}

\subsubsection{Binary LJ Cluster}

Next, we examine the binary A$_1$B$_{12}$ cluster, for which geometry-optimized structures are available as benchmarks. \cite{mravlak_structure_2016}
We reuse the shifted, truncated LJ potential from Section~\ref{demo-lj6}, assigning distinct well depths to represent the two components:
$\varepsilon_{\mathrm{AA}}=0.1$~eV for A--A interactions,
$\varepsilon_{\mathrm{BB}}=0.05$~eV for B--B interactions, and
$\varepsilon_{\mathrm{AB}}=\sqrt{\varepsilon_{\mathrm{AA}}\varepsilon_{\mathrm{BB}}}\approx0.07$~eV (Berthelot rule) for A--B interactions.
To isolate the effect of interaction-energy asymmetry on the \CV curve, we set the particle sizes equal ($\sigma_{\mathrm{A}}=\sigma_{\mathrm{B}}=2.5$~\AA).
The particles are placed in a non-periodic cubic box of edge length 15.41~\AA, corresponding to a number density of $5.56\times10^{-2}~\sigma^{-3}$.
Simulations were run as follows:
NS simulations used 960 walkers with 400 decorrelation steps, a significant increase over LJ$_6$ to account for the more than doubled degrees of freedom from 13 particles;
MC was cooled from 400 to 100~K in 25~K increments; and
WL sampled energies from 0 to -2.44~eV, slightly above the ground-state energy (-2.45~eV).

As with LJ$_6$, the A$_1$B$_{12}$ cluster exhibits a gas--solid transition at higher temperatures and a solid--solid transition at lower temperatures.
Unlike LJ$_6$'s octahedral ground state, however, the low-temperature structure adopts an A--B icosahedral core--shell configuration. \cite{mravlak_structure_2016}
Figure~\ref{fig:LJ_A1B12} shows that all three methods reproduce these features, with closely matching transition temperatures in units of $k_{\mathrm{B}}T/\varepsilon$:
0.276 (NS), 0.280 (MC), and 0.275 (WL) for the gas--solid transition, and
0.169 (NS), 0.172 (MC), and 0.167 (WL) for the solid--solid transition.
Compared to LJ$_6$, the A$_1$B$_{12}$ system shows a slightly lower gas--solid transition temperature and a markedly higher, sharper solid--solid transition. 
{%\color{blue}
The ground-state structure corresponds to the minimum-energy configuration in the binary Lennard-Jones database, \cite{mravlak_structure_2016} and the double-peaked \CV curve and associated phase changes are consistent with descriptions from early work using argon- and neon-like Lennard-Jones parameters.\cite{sabo2004phase}
}

\subsubsection{LJ Surfaces}

\begin{figure*}[tb]
\centering
\begin{mdframed}
\begin{singlespace} 
\begin{minipage}{1\textwidth}
\begin{minted}[linenos,breaklines,fontsize=\scriptsize]{julia}
# Read walkers from a file; periodic boundary conditions set to (true, true, false)
ats = read_walkers("slab.extxyz", pbc="TTF") # 320 configurations stored in this file -> 320 walkers

# Split each walker into two components: 80 frozen particles (surface) and 4 free particles (adsorbates)
walkers = [AtomWalker{2}(at.configuration; list_num_par=[80,4], frozen=Bool[1,0]) for at in ats]

# Create an LJ potential
lj = LJParameters(epsilon=0.1, sigma=2.5, cutoff=4.0, shift=true)

# Nested sampling
ns_energies, ns_ls, ns_params = nested_sampling(LJAtomWalkers(walkers, lj), NestedSamplingParameters(mc_steps=200, step_size=0.1), 50_000, MCRandomWalkClone(), SaveEveryN(n_traj=10))

# Metropolis Monte Carlo using a temperature grid from 2000 K down to 100 K, in 100 K steps
mc_energies, mc_ls, mc_cvs, acceptance_rates = monte_carlo_sampling(walkers[1], lj, MetropolisMCParameters(collect(2000.0:-100:100), equilibrium_steps=5_000_000, sampling_steps=5_000_000, step_size=1.0))

# Wang-Landau sampling
wl_energies, wl_ls, wl_params, S, H = wang_landau(walkers[1], lj, WangLandauParameters(num_steps=10_000, energy_min = -57.06, energy_max=-50.0, num_energy_bins=1_000, step_size=1.0, f_min=1.00001))
\end{minted}
\end{minipage}
\end{singlespace}
\centering
\includegraphics[width=0.8\linewidth]{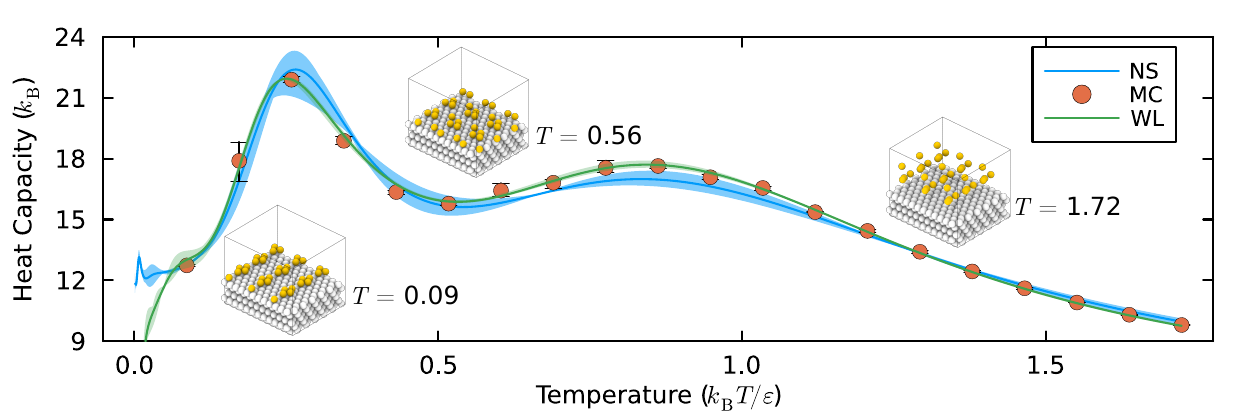}
\end{mdframed}
\caption{
Julia code and results for an LJ(111) surface with quarter coverage ($\theta=1/4$), showing the constant-volume heat capacity computed using nested sampling (NS), Metropolis Monte Carlo (MC), and Wang--Landau (WL) sampling.
Each method was repeated three times; NS and WL are shown as lines with shaded ranges (min--max), and MC as markers with error bars (min--max).
Insets illustrate temperature-dependent structural transitions, with adsorbates reorganizing on the surface.
Representative structures are taken from NS and corroborated with MC; for visualization, the $4\times4$ per-monolayer unit cell is replicated twice in the lateral dimensions.
}
\label{fig:LJ_surf}
\end{figure*}

Finally, we examine the (111) surface of a face-centered cubic LJ solid, building on previous NS benchmarks. \cite{yang_surface_2024, chatbipho_adsorbate_2025}
We model a five-layer slab containing 80 fixed particles (16 per layer) and four mobile adsorbates, corresponding to a maximum coverage of $\theta=1/4$.
The simulation box has periodic boundaries in the in-plane directions ($a=11.2$~\AA, $b=9.73$~\AA) and a non-periodic upper boundary at $c=29.15$~\AA\, where simple reflection is applied if a particle moves beyond this limit.
Sampling parameters are as follows:
NS ran for 50,000 steps with 320 walkers;
MC cooled from 2000 to 100~K in 100~K increments; and
WL sampled energies from -50.0 to -57.06~eV, slightly above the ground-state energy (-57.07~eV) (Figure~\ref{fig:LJ_surf}).

This LJ(111) quarter-coverage (q.c.) system is expected to display
a broad \CV peak at $0.86~k_{\mathrm{B}}T/\varepsilon$ (surface condensation) and
a sharper peak at $0.25~k_{\mathrm{B}}T/\varepsilon$ (two-dimensional adsorbate ordering). \cite{yang_surface_2024}
Figure~\ref{fig:LJ_surf} shows that all three methods yield consistent transition temperatures in units of $k_{\mathrm{B}}T/\varepsilon$:
0.833 (NS), 0.862 (MC), and 0.849 (WL) for the condensation transition; and
0.265 (NS), 0.259 (MC), and 0.247 (WL) for the ordering transition.
\textblue{These findings are in exact agreement with our earlier study employing an independent nested sampling implementation.} \cite{yang_surface_2024}

\subsubsection{Discussion of Atomistic Demonstrations}

For the simple atomistic systems examined, NS, MC, and WL produced consistent configurational statistics and thermodynamic properties.
The agreement among the three methods, and with published benchmarks, \cite{partay_efficient_2010, mravlak_structure_2016, yang_surface_2024, maillard_nested_2025} confirms that they are correctly implemented within \freebird.
This unified framework allows verification of sampling outputs by cross-checking across methods using a consistent call structure, making it straightforward for users to apply multiple algorithms to their systems of interest.

\begin{table}[tb]
\caption{\label{tab:eval}
The number of energy evaluations used in each of nested sampling (NS), Metropolis Monte Carlo (MC), and Wang--Landau (WL) sampling, in millions.
For MC, this value applies to each temperature individually.
}
% \begin{ruledtabular}
\begin{tabular}{
    lccc
}
    \hline
    System           & \text{NS} & \text{MC (at each $T$)} & \text{WL} \\ \hline
    LJ$_6$           &   5.0     &  0.2                    &  247.9    \\
    LJ A$_1$B$_{12}$ & 200.0     & 15.0                    & 6735.8    \\
    LJ(111) q.c.     &  12.5     & 10.0                    &  330.5    \\ \hline
\end{tabular}
% \end{ruledtabular}
\end{table}

Because all sampling methods use the same random-walk and energy-evaluation routines, the total number of energy evaluations provides a consistent probe of computational cost (Table~\ref{tab:eval}).
\textblue{The reported values range from a few million to several billion evaluations, illustrating the order-of-magnitude differences in computational effort between algorithms.}
Note that these costs vary with both system complexity and algorithm choice.
Table~\ref{tab:eval} is not intended as a direct performance comparison, since the methods differ in their intrinsic overheads (e.g., number of walkers for NS, temperature grids for MC, and number of bins plus flatness criterion for WL), and each can be further tuned to balance accuracy and cost.
The reported values serve only as references for results obtained using straightforward, representative parameter settings for each algorithm.

In LJ$_6$, NS required 25,000 iterations with 200 decorrelation steps each, totaling 5 million evaluations.
MC ran 4 million evaluations across 20 temperatures.
WL, which flattened a 1000-bin histogram to $f<1.00001$ over 16 iterations using 24,787 batches of 10,000 walks, required nearly 248 million evaluations, which is an order of magnitude higher than NS or MC, and produced a well-converged, low-variance heat-capacity trace.

In the binary LJ A$_1$B$_{12}$ cluster, configurational complexity increases due to the larger number of particles and the presence of a second species, introducing compositional disorder.
NS used 960 walkers and nearly 500,000 iterations (at 400 steps each), totaling 200 million evaluations.
MC performed 15 million evaluations per temperature, while WL required over 6.7 billion evaluations to flatten a 1000-bin histogram to $f<1.00001$.

In the LJ(111) slab with adsorbates, the number of mobile particles is small, but the fixed substrate creates a complex PES with many near-degenerate configurations.
NS required 320 walkers (more than double that of LJ$_6$) and over twice the number of energy evaluations.
MC sampling increased 50-fold (from 0.2 to 10 million evaluations per temperature), especially near phase transitions.
WL required a similar effort as for LJ$_6$ but required significantly more energy evaluations than the other two sampling methods.

Although direct comparisons are challenging due to differing objectives and strategies, the sampling methods exhibit distinct trade-offs.
NS is well-suited for generating configurations that span a wide energy range, from gas-like states to ground-state structures, making it effective for identifying candidate transition regions.
Its stepwise nature limits parallelism compared with approaches such as parallel tempering, \cite{swendsen_replica_1986} but its walker-removal rate can be tuned to emphasize either exploratory breadth or resolution. \cite{martiniani_superposition_2014}
This flexibility makes NS a strong first-pass strategy for locating regions of interest for subsequent refinement with other algorithms.
MC is advantageous when properties at specific temperatures are of primary interest.
WL, though computationally demanding for continuous systems, \cite{zhou_understanding_2005} can yield low-variance estimates once convergence is reached, due to its comprehensive sampling of configurations finely spaced in energy.
In practice, the methods can complement each other:
NS can identify temperature ranges for MC and guide energy bounds for WL, while
MC energy histograms can inform WL binning resolution within those bounds.
As a modular toolbox, \freebird supports not only method selection but also coordination across algorithms.
The three demonstrations underscore the framework's flexibility, accommodating varying numbers of components, interaction schemes, and physical constraints.
As the codebase evolves, it will continue to support increasingly complex and realistic systems.

\subsection{Lattice Systems}

To demonstrate \freebird's applicability to models with discretely positioned particles, we consider two representative lattice systems:
(1) a two-dimensional square lattice with partially occupied sites, and
(2) a three-dimensional primitive cubic lattice that models adsorption and desorption across layers.

\subsubsection{Two-Dimensional Lattice Models\label{demo-2d-lattice-models}}

\begin{figure*}[tb]
\centering
\begin{mdframed}
\begin{singlespace} 
\begin{minipage}{1\textwidth}
\begin{minted}[linenos,breaklines,fontsize=\scriptsize]{julia}
# 2D lattice example
# Initial single-component square lattice (4×4×1 by default) with sites [1,2,3,4] occupied
initial_lattice = SLattice{SquareLattice}(components=[[1,2,3,4]])

# Hamiltonian: adsorption energy = -0.04 eV, nearest-neighbor energy = -0.01 eV, next-nearest-neighbor energy = -0.0025 eV
h = GenericLatticeHamiltonian(-0.04, [-0.01, -0.0025], u"eV")

# Exact enumeration
exact_energies, exact_configs = exact_enumeration(initial_lattice, h)

# Nested sampling
walkers = [LatticeWalker(generate_random_new_lattice_sample!(initial_lattice)) for _ in 1:1000]
ns_energies, ns_ls, ns_params = nested_sampling(LatticeGasWalkers(walkers, h), LatticeNestedSamplingParameters(), 10_000, MCRejectionSampling(), SaveEveryN())

# Metropolis Monte Carlo using a temperature grid from 200 K down to 10 K, in 10 K steps
mc_energies, mc_configs, mc_cvs, acceptance_rates = monte_carlo_sampling(initial_lattice, h, MetropolisMCParameters(collect(200.0:-10:10), equilibrium_steps=25_000, sampling_steps=25_000))

# Wang-Landau sampling
wl_energies, wl_configs, wl_params, S, H = wang_landau(initial_lattice, h, WangLandauParameters(energy_min=-0.20625, energy_max=-0.15875))

# A 3D lattice with three layers can be constructed by setting supercell_dimensions=(4,4,3). The sampling lines above can then be rerun (with more NS iterations/MC steps) to obtain results:
initial_lattice = SLattice{SquareLattice}(components=[[1,2,3,4]], supercell_dimensions=(4,4,3), adsorptions=collect(1:16))
\end{minted}
\end{minipage}
\end{singlespace}
% \vspace{1em}
\centering
\includegraphics[width=0.425\linewidth]{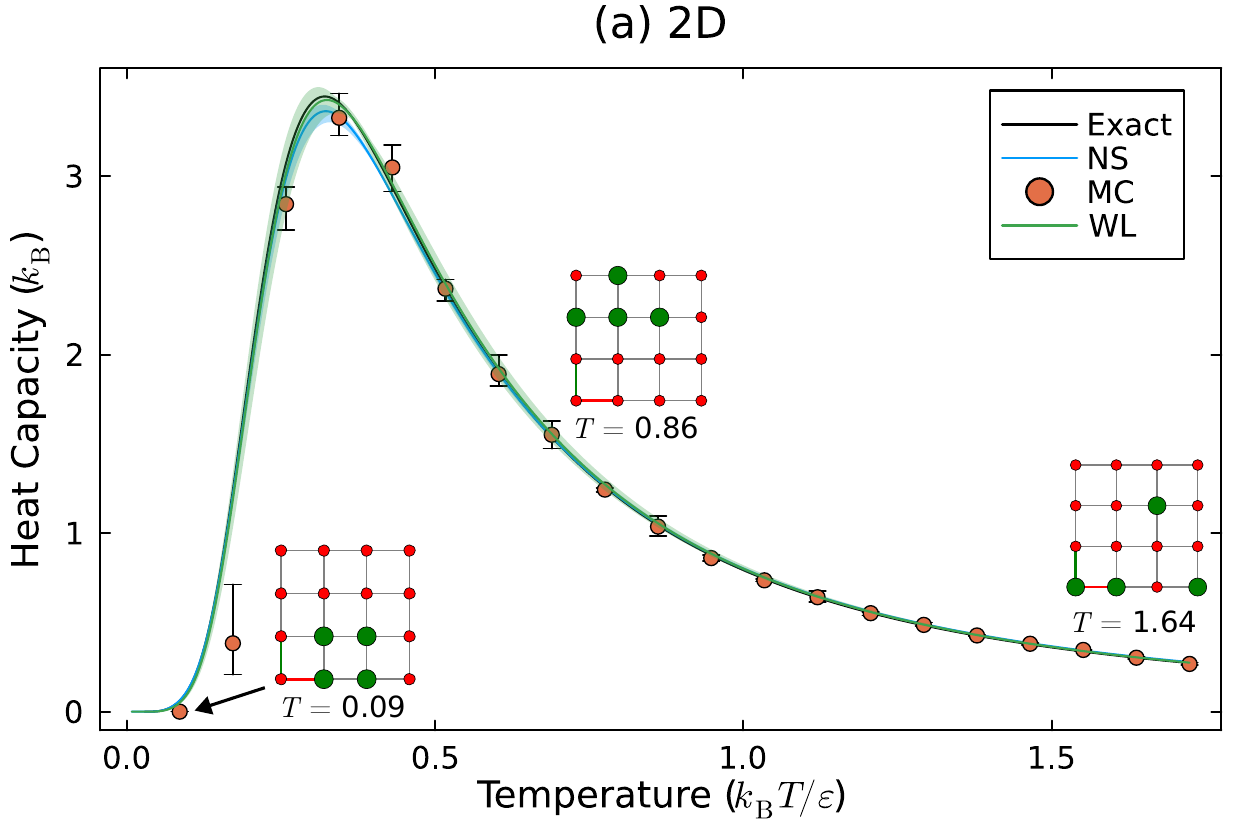}
\includegraphics[width=0.425\linewidth]{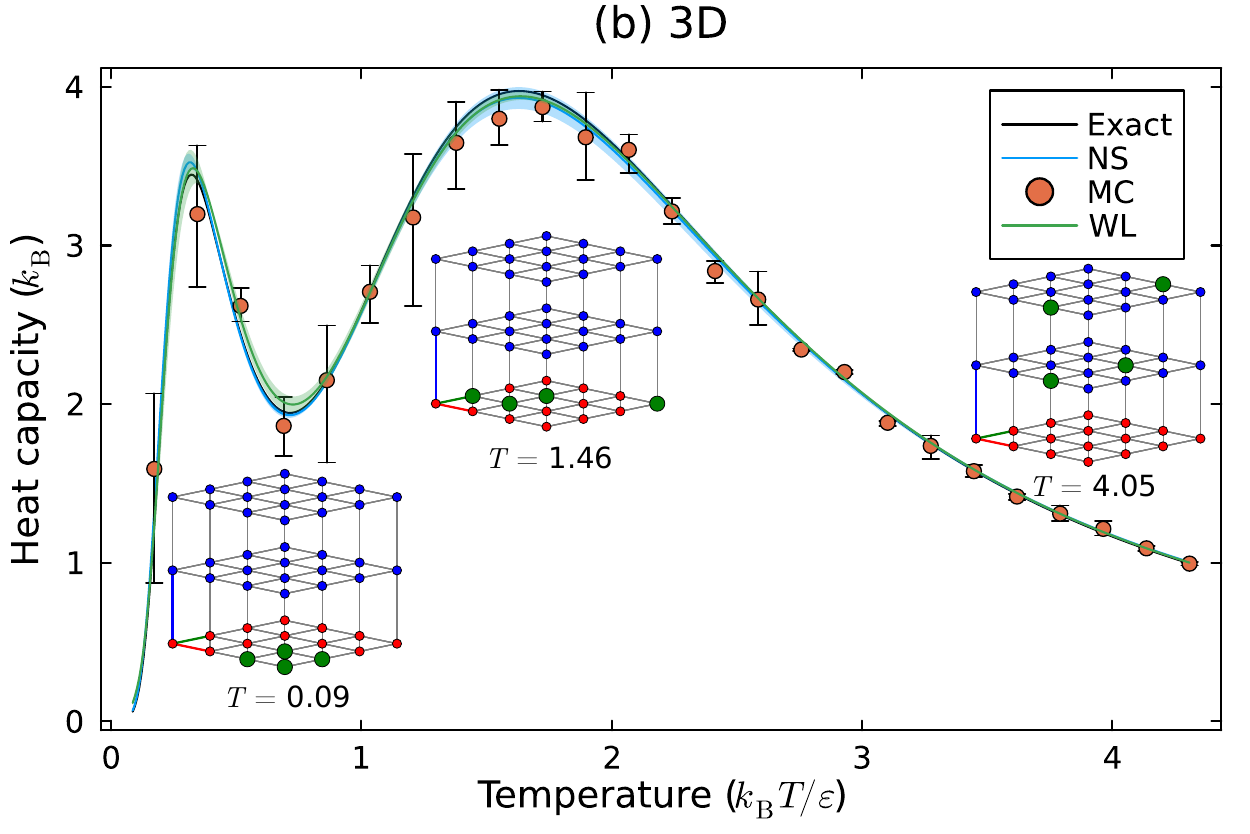}
\end{mdframed}
\caption{
Julia code and results for lattice models with discrete occupancy:
(a) a two-dimensional $4\times4\times1$ lattice and
(b) a three-dimensional $4\times4\times3$ lattice, both with fractional coverage $\theta=4/16$.
The constant-volume heat capacity is computed using nested sampling (NS), Metropolis Monte Carlo (MC), and Wang--Landau (WL) sampling.
Each method was repeated three times; NS and WL are shown as lines with shaded ranges (min--max), and MC as markers with error bars (min--max).
Exact results were obtained by full enumeration of all configurations.
Insets show the configurational change from high to low temperature, with blue circles indicating lattice sites, red circles indicating adsorption sites, and green circles indicating occupied sites.
}
\label{fig:lattice}
\end{figure*}

We begin with a two-dimensional $4\times4$ square lattice containing four occupied sites, corresponding to a coverage of $\theta=1/4$.
This system serves as a coarse-grained analog of LJ(100) q.c.\ and admits exact thermodynamic results via enumeration.
The lattice Hamiltonian is defined as
\begin{equation}
    H = N \varepsilon_{\mathrm{ads}} + \sum_{\mathrm{nn}} \varepsilon_{\mathrm{nn}} + \sum_{\mathrm{nnn}} \varepsilon_{\mathrm{nnn}} + \cdots
\end{equation}
where $N$ is the number of occupied sites; $\varepsilon_{\mathrm{ads}}$ is the adsorption energy per occupied site; and $\varepsilon_{\mathrm{nn}}$ and $\varepsilon_{\mathrm{nnn}}$ are the interaction energies between nearest-neighbor (nn) and next-nearest-neighbor (nnn) pairs of occupied sites, respectively.
We set $\varepsilon_{\mathrm{nn}}=-0.01$~eV to model exothermic pairing between nearest neighbors and $\varepsilon_{\mathrm{ads}}=4\varepsilon_{\mathrm{nn}}=-0.04$~eV to account for adsorption at the centers of four implicit surface particles.
To mimic the distance decay of an LJ potential, we use $\varepsilon_{\mathrm{nnn}}=\varepsilon_{\mathrm{nn}}/4=-0.0025$~eV, reflecting the approximate quartering of LJ interaction strength from $r_{\mathrm{min}}=2^{1/6}\sigma$ to $\sqrt{2}r_{\mathrm{min}}$.
We perform
NS with 1,000 walkers and a constant walker-removal rate $C=1$, using \codeword{MCRejectionSampling()} to generate new configurations and progressing only when a lattice configuration of equal or lower energy is found;
MC with a temperature increment of 10~K (from 200~K to 10~K), and
WL sampling using 100 energy bins spanning $\left[-0.20625,-0.15875\right]$~eV.
This range extends slightly beyond the exact bounds, minimum energy -0.205~eV and maximum -0.16~eV, by half the smallest interaction magnitude ($|\varepsilon_{\mathrm{nnn}}|/2=0.00125$~eV).
Exact enumeration of all 1,820 configurations reveals a single order--disorder transition at $0.321~k_{\mathrm{B}}T/\varepsilon$, favoring a square ground state.
As shown in Figure~\ref{fig:lattice}, all three methods detect this transition, yielding consistent transition temperatures in units of $k_{\mathrm{B}}T/\varepsilon$:
0.323 (NS), 0.345 (MC), and 0.325 (WL).

\subsubsection{Three-Dimensional Lattice Models}

We next examine a three-dimensional $4\times4\times3$ primitive cubic lattice with four occupied sites.
The bottom layer represents an adsorbent surface, while the upper two layers act as a fluid phase into which particles may desorb, emulating surface condensation.
The same lattice Hamiltonian is used as in Section~\ref{demo-2d-lattice-models}, except that $\varepsilon_{\mathrm{ads}}=0$~eV is assigned to occupied sites in the fluid layers to isolate the effect of fluid-phase introduction.
Only nn and nnn interactions are considered.
Simulation parameters match those of the two-dimensional model.
Exact enumeration of the full configuration space (194,580 configurations) again reveals an order--disorder transition at lower temperatures ($0.319~k_{\mathrm{B}}T/\varepsilon$, nearly identical to the $0.321~k_{\mathrm{B}}T/\varepsilon$ transition in the 2D case), favoring the square ground state.
In contrast to the two-dimensional system, however, the three-layer model also shows a broad \CV peak at higher temperatures ($1.629~k_{\mathrm{B}}T/\varepsilon$), corresponding to a surface condensation transition.
As shown in Figure~\ref{fig:lattice}, all three methods capture both features with closely matching transition temperatures:
0.319 (NS), 0.259 (MC), and 0.327 (WL) for the order--disorder transition; and
1.629 (NS), 1.637 (MC), and 1.637 (WL) for the surface condensation transition.
\textblue{Compared to the 2D case, the computational effort for all methods was increased to ensure convergence. The number of NS iterations was raised from 10,000 to 30,000; MC equilibration and sampling steps at each temperature were increased from 25,000 to 100,000; and WL retained the same number of energy bins (since the potential-energy landscape is unchanged), but the number of MC proposals per attempt to reduce $f$ was increased from 100 to 1,000 to account for the larger number of microstates. These settings were used to generate the results shown in Figure~\ref{fig:lattice}(b).}

\subsubsection{Discussion of Lattice Demonstrations}

Lattice models can greatly reduce computational cost while still capturing the essential surface phase transitions observed in atomistic systems.
For small lattices, exact enumeration is tractable:
the $4\times4$ 2D model comprises 1,820 configurations, which can be exhaustively sampled in under one second on a single CPU core with a single thread, whereas
the 3D model with three layers and four adsorbates contains 194,580 configurations and can be enumerated in about 30 seconds under the same conditions.
WL sampling performs particularly well on these discrete systems, benefiting from the finite number of energy states and their known spacing.
MC results are generally consistent with WL but show greater uncertainty near phase transitions, especially at low temperatures, where dense phases dominate and single-particle MC moves struggle to interconvert between competing low-energy configurations.

NS poses additional challenges.
In lattice models, discrete energy degeneracies produce plateaus in the NS likelihood--energy profile. \cite{murray_nested_2005, schittenhelm_nested_2021, fowlie_nested_2021}
To break these degeneracies, we follow the approach of Murray et al.\ and recent \texttt{BraWl} implementations, adding to each configuration's energy a uniformly distributed perturbation much smaller than the typical level spacing (e.g., $<10^{-12}$~eV). \cite{murray_nested_2005, naguszewski_brawl_2025}
With single-site random-walk moves, lattice NS shows sharply reduced acceptance rates as the energy ceiling approaches the ground-state energy.
\textblue{As shown in Figure~\ref{fig:lattice}, 10,000 NS samples are required for the 2D lattice to establish the correct distribution for constructing an accurate partition function, exceeding the total number of microstates in the system.}
Introducing cluster or collective moves, or increasing the live-set size, can restore sampling efficiency. \cite{swendsen_nonuniversal_1987}
Even so, WL remains the most efficient method for these systems.

Collectively, these demonstrations underscore \freebird's versatility in implementing and contrasting diverse sampling strategies within a single lattice system, thereby facilitating both performance benchmarking and thermodynamic cross-validation.

\section{Conclusions and Outlook\label{sec:conclusions}}

In this work, we introduced \freebird, a comprehensive, flexible, and extensible toolbox for modeling solid interfaces, implemented in the modern programming language Julia.
\freebird provides multiple sampling methods (Metropolis sampling, Wang--Landau sampling, and nested sampling) within a unified framework that employs a common structure-handling system and energy calculators across all methods.
Its flexibility arises from the use of abstract data types, which can be easily customized through convenient constructor functions.
These data types also facilitate extending the code to new classes of systems beyond the already general atomistic continuous systems and discretized lattices.
\freebird's extensibility further enables the incorporation of additional sampling methods (e.g., replica exchange) using the existing data structures, as well as the implementation of alternative interaction models beyond simple pairwise potentials.

Julia, as a high-level, just-in-time compiled language, significantly lowers the barrier to developing complete, high-performance computational chemistry packages. \cite{pulsipher2025optimization}
The availability of high-quality standard libraries and community packages in the Julia ecosystem, particularly those from JuliaMolSim, has been instrumental in the rapid development of \freebird.
Examples include \codeword{AtomsBase.jl} \cite{atomsbase_jl} for handling atomic and cell properties, \codeword{AtomsIO.jl} \cite{atomsio_jl} and \codeword{ExtXYZ.jl} \cite{extxyz_jl} for reading and writing atomic structures, and the Julia standard libraries \codeword{Threads} and \codeword{Distributed} for straightforward parallelization, which substantially improves \freebird's performance and enables execution on HPC systems.
Contributors to this open-source package span a range of career stages, from undergraduate students to established computational scientists, underscoring its accessibility to both users and developers.
Finally, \freebird exemplifies the adoption of modern software engineering practices to enhance productivity, sustainability, and reproducibility, serving as a template for the development of contemporary scientific software.

Future developments of \freebird can be categorized into algorithmic and implementation enhancements.
Algorithmically, integrating the existing sampling methods with ensembles beyond the canonical, such as the semi-grand-canonical ensemble (to allow composition changes) and the grand-canonical ensemble (for chemical potential control), will expand the range of systems and properties relevant to materials discovery.
Additionally, while the current move sets are adequate for many systems, more sophisticated proposals (e.g., cluster or collective moves) are often necessary near critical points or in systems with slow dynamics to ensure efficient sampling. \cite{swendsen_nonuniversal_1987, wolff_collective_1989}
From an implementation perspective, enabling \freebird for efficient GPU computing is essential for scaling to exascale platforms.
This capability can be combined with machine-learning interatomic potentials for energy evaluations, which naturally exploit GPU acceleration.
Overall, \freebird functions not only as a performant tool for sampling computations, but also as an integrative interface linking diverse components of molecular simulation, and as a versatile platform for advancing computational chemistry and materials science.

\begin{acknowledgement}

% Please use ``The authors thank \ldots'' rather than ``The
% authors would like to thank \ldots''.

% The author thanks Mats Dahlgren for version one of \textsf{achemso},
% and Donald Arseneau for the code taken from \textsf{cite} to move
% citations after punctuation. Many users have provided feedback on the
% class, which is reflected in all of the different demonstrations
% shown in this document.

RBW acknowledges support from the National Science Foundation under Grant No.~2305155.
A portion of this research was conducted as part of a user project at the Center for Nanophase Materials Sciences (CNMS), which is a US Department of Energy, Office of Science User Facility at Oak Ridge National Laboratory.
This research used resources of the National Energy Research Scientific Computing Center (NERSC), a Department of Energy User Facility.

\end{acknowledgement}

\begin{suppinfo}

Data Availability:
The computational package \freebird is an open-source project publicly available at \url{https://github.com/wexlergroup/FreeBird.jl}.
The data that support the findings of this study, as well as the \freebird input files for generating the data, are openly available at \url{https://github.com/wexlergroup/FreeBird.jl-paper-data-2025}.

\end{suppinfo}

\bibliography{references-no-notes,packages,additions}

\end{document}